\documentclass[balance=false, sigconf]{acmart}

\usepackage{pifont}
\usepackage{xspace}
\usepackage{multirow} 
\usepackage{soul}
\usepackage{changepage,threeparttable} 
\usepackage{marginnote}
\usepackage{caption}
\usepackage{algorithm}
\usepackage{algpseudocode}
\usepackage{xcolor}
\usepackage{listings}

\definecolor{BoxGray}{gray}{0.93}
\long\def\finding#1{\hspace{-0.5cm} \colorbox{BoxGray}{\fbox{\parbox{0.95\columnwidth}{\emph{#1}}}}}

\newcommand{\ea}{{et al.}\xspace}
\newcommand{\thead}[1]{\textbf{\textit{#1}}}

\graphicspath{{./figures/}}

\definecolor{codegreen}{rgb}{0,0.6,0}
\definecolor{backcolour}{rgb}{0.95,0.95,0.92}
\definecolor{LightGray}{gray}{0.9}
\definecolor{Amber}{rgb}{1.0, 0.75, 0.0}

\AtBeginDocument{%
  \providecommand\BibTeX{{%
    \normalfont B\kern-0.5em{\scshape i\kern-0.25em b}\kern-0.8em\TeX}}}

\setcopyright{acmcopyright}
\copyrightyear{2024}
\acmYear{2024}
\acmDOI{XXXXXXX.XXXXXXX}

\acmConference[Published in MSR '24]{the 21st International Conference on Mining Software Repositories}{April 2024}{Lisbon, Portugal}
\acmPrice{15.00}
\acmISBN{978-1-4503-XXXX-X/18/06}

\begin{document}

\title{Analyzing and Mitigating (with LLMs) the Security Misconfigurations of Helm Charts from Artifact Hub}

\author{Francesco Minna}
\email{f.minna@vu.nl}
\orcid{0002-3018-044X}
\affiliation{%
  \institution{Vrije Universiteit Amsterdam}
  \country{Netherlands}
}

\author{Fabio Massacci}
\email{fabio.massacci@ieee.org}
\orcid{0002-1091-8486}
\affiliation{%
  \institution{Vrije Universiteit Amsterdam and the University of Trento}
  \country{Netherlands, Italy}
}

\author{Katja Tuma}
\email{k.tuma@vu.nl}
\orcid{0001-7189-2817}
\affiliation{%
  \institution{Vrije Universiteit Amsterdam}
  \country{Netherlands}
}

\renewcommand{\shortauthors}{Minna, et al.}

\begin{abstract}
\textbf{Background}: Helm is a package manager that allows defining, installing, and upgrading applications with Kubernetes (K8s), a popular container orchestration platform.
A Helm chart is a collection of files describing all dependencies, resources, and parameters required for deploying an application within a K8s cluster.

\noindent
\textbf{Objective}: The goal of this study is to mine and empirically evaluate the security of Helm charts, comparing the performance of existing tools in terms of misconfigurations reported by policies available by default, and measure to what extent LLMs could be used for removing misconfigurations.
We also want to investigate whether there are false positives in both the LLM refactorings and the tool outputs.

\noindent
\textbf{Method}: We propose a pipeline to mine Helm charts from Artifact Hub, a popular centralized repository, and analyze them using state-of-the-art open-source tools, such as Checkov and KICS. 
First, such a pipeline will run several chart analyzers and identify the common and unique misconfigurations reported by each tool. 
Secondly, it will use LLMs to suggest mitigation for each misconfiguration.
Finally, the chart refactoring previously generated will be analyzed again by the same tools to see whether it satisfies the tool's policies.
At the same time, we will also perform a manual analysis on a subset of charts to evaluate whether there are false positive misconfigurations from the tool's reporting and in the LLM refactoring.
\end{abstract}



\maketitle

\section{Introduction}

Kubernetes (K8s) has become the de-facto standard tool to orchestrate container-based applications in the cloud, reported to be used by 64\% of organizations in production environments from the 2022 Cloud Native Computing Foundation (CNCF) annual survey~\cite{cncf-2022-report}.
%
Helm is a package manager which is used to help deploy applications on K8s clusters.
%
It allows defining applications as charts, a set of folders and files that bundle all resources and dependencies needed by an application, reducing the need to manually write and manage K8s YAML files. 

These charts often contain different types of misconfigurations~\cite{Bridgecrew,3579639}, and several static analysis tools have been developed to detect them, such as Checkov~\cite{checkov}, Datree~\cite{datree}, and KICS~\cite{kics}, to name a few.
Based on rule-based policies, these tools analyze Helm charts and K8s YAML files and highlight security risks, such as missing memory limits or an over-privileged security context for a container.
Such tools can detect configuration lines in a YAML file, such as the \texttt{allowPrivilegeEscalation: true} option, that would allow a container to escalate privileges during execution, which is considered a bad practice, according to the K8s CIS Benchmarks~\footnote{\url{https://www.cisecurity.org/cis-benchmarks}}.
However, even if some of these misconfigurations may seem trivial to detect, they are still very present in the wild; additionally, other options, such as wrong network policies, may not be trivial to detect.

For example, a blog post by Checkov developers describes the number and type of misconfigurations reported by Checkov on several charts~\cite{Bridgecrew}.
For such misconfigurations, there exist several rule-based policies (e.g., ``Prevent containers from escalating privileges'' --- \texttt{allow Privilege Escalation: false}), that check whether a YAML key is present or not, and eventually its value.

Existing tools have different policies and different configurations may satisfy the same policy; however, most rule-based tools do not suggest or implement any mitigations for the corresponding misconfigurations.
On the other hand, Large Language Models (LLM), such as Google Gemini~\cite{gemini} and ChatGPT~\cite{chatgpt}, allow, among other things, to write, understand, and refactor source code.
Therefore, we are interested in evaluating whether LLMs could be used to refactor a K8s deployment file by implementing new safer configurations.
Since LLMs can generate wrong and inconsistent results, for instance, due to flawed training data sources~\cite{huang2023survey}, we are also interested in measuring their reliability to determine whether they can be used to generate secure cloud configurations.
Indeed, a recent study has found that LMM-assisted
users produced critical security bugs at a rate no greater than
10\% more compared to users coding without LLMs~\cite{sandoval2023lost}.
%

The goal of this study is to mine and empirically evaluate the security of Helm charts, in terms of misconfigurations, compare the performance of existing tools, and measure to what extent LLMs could be used for removing misconfigurations.
In doing so, we plan to investigate what is the general status of charts available on Artifact Hub, for example, in terms of the number of misconfigurations reported.
We are also interested in what policies are available by default in each tool, and whether there are unique policies or, in other words, if certain misconfigurations can only be detected by one tool.
At the same time, we also want to investigate if and how Large Language Models (LLM) can be used to remove misconfigurations, and whether the implemented mitigations will satisfy the tool policies.
By doing so, this study will shed light on the security of Helm charts, existing tools, and whether LLMs can improve it.

The rest of this registered report is organized as follows. 
In the next section~\S\ref{sec:related_work}, we provide some background and related work about Helm Charts security and configuration repair. 
In~\S\ref{sec:research_questions}, we present our research questions, and in~\S\ref{sec:execution_plan} we present our execution plan to answer such questions. 
Finally, in~\S\ref{sec:preliminary_study} we present some findings from our preliminary study, concluding with the threats to validity~\S\ref{sec:threats_to_validity}.

\section{Background and Related Work}
\label{sec:related_work}

This section presents the background on the Helm package manager, Helm charts, and tools that can analyze such charts.

\subsection{Infrastructure as Code and Helm Charts} 

Infrastructure as Code (IaC) enables provisioning, managing, and configuring infrastructure resources (such as virtual machines and containers) using configuration files, rather than manual processes. 
This allows for easy versioning, sharing, and scaling.
Previous work has already investigated the security of IaC scripts, for example, to identify defects and privacy violations~\cite{31834402} and to predict and identify defects in open-source project commits~\cite{23183440}.
Rahman~\ea\cite{8812041} evaluated $1,726$ IaC scripts to identify the presence of seven security smells --- weak code patterns that can lead to vulnerabilities, e.g., suspicious comments and weak cryptographic algorithms, the frequency, and mitigation lifetime. The same authors found $9,175$ hard-coded passwords in IaC scripts~\cite{9388795}.


Kubernetes (K8s) is a container orchestration engine that enables the efficient management, deployment, and scaling of hundreds of containers using YAML configuration files.
Helm is a package manager for K8s that allows to define, manage, share, version, and deploy applications as a set of directories and YAML files, called charts.
Artifact Hub~\cite{artifacthub} is a repository where developers and organizations can upload, share, and download Helm chart artifacts; in other words, it is the equivalent of Docker Hub for container images, but for Helm Charts.

\subsection{Security Issues in Helm Charts}

As pointed out by previous work, K8s configuration files or manifests contain misconfigurations and security defects. For example, Bose~\ea\cite{9476056} and Rahman~\ea~\cite{3579639} investigated the type and frequency of security defects (e.g., buffer overflow and denial of service) that appear in K8s manifest from open-source projects.
%
%
Shamim~\ea~\cite{9230176} surveyed $104$ grey literate source and suggested $11$ security best practices, such as CPU and memory request limits, network policies, and avoiding the use of default namespaces.
Finally, Blaise~\ea~\cite{9860863} used topological graphs to represent Helm Charts and scored each deployment resource based on best practices, container image vulnerabilities, and potential attack paths.

Besides academic contributions, there also exist several tools that can detect misconfigurations, security risks, and best-practice violations in such charts before deployment.
Examples of such tools are Checkov, Datree, KICS, and Terrascan.
To the best of our knowledge, a study investigating the characteristics and performance of such tools is still missing.

\subsection{Misconfiguration Repair}

Automatic Program Repair (APR) techniques aim to automatically identify patches for a given error or bug, minimizing human intervention, which is lately receiving a lot of attention.
%
%
Pinconschi~\ea\cite{3533767} proposed MAESTRO, an APR pipeline to test several vulnerability benchmarks (e.g., Defects4J and CB-repair) and APR techniques (e.g., GenProg and MUT-APR) for C and Java programs; they evaluated the pipeline on ten projects, evaluating correctness and time overhead.
Tjiong~\ea\cite{9970223} proposed F1X, an APR tool to fix security vulnerabilities found by OSS-Fuzz, a fuzzing framework to find vulnerabilities in open-source projects; they evaluated the tool on $240$ vulnerabilities found by OSS-Fuzz in five C open-source projects, measuring correctness compared to OSS-Fuzz suggestions.
Fu~\ea\cite{35402508} proposed VulRepair, a T5-based (a Transformer-based Neural Machine Translation --- NMT) vulnerability repair approach; they evaluated the approach on more than $8,000$ vulnerability fixes from $1,754$ open-source projects, measuring accuracy and fix prediction.

Large language models (LLMs) are a type of artificial intelligence (AI) based on deep learning algorithms, trained on a massive amount of text data, that can perform several natural language processing (NLP) tasks.
For example, they can generate new text, translate, understand contexts, and answer questions; in other words, they output the most likely next word based on the given input.
Google Gemini and ChatGPT are two common examples of LLMs.

LLMs have already been investigated in previous work, discussing the advantages and disadvantages of usages for software engineering practices~\cite{fan2023large}, presenting a prompt patterns catalog for software engineering tasks~\cite{white2023chatgpt}, and evaluating the code generation capabilities of both ChatGPT and Copilot~\cite{Hansson1764568}.
Biswas~\ea\cite{Biswas2023} discussed ChatGPT software programming capabilities; among other results, they concluded that ChatGPT can perform code completion, correction, and refactoring while providing users with explanations.  
Sokolowski~\ea\cite{10092598} proposed an automated testing tool to fuzz Pulumi TypeScript IaC programs for reliability, and testing configurations for termination, correctness, and compliance.
However, besides the detection and refactoring advantages of using LLMs, there is also the concern of recommending or generating insecure code, as shown by Pearce~\ea\cite{9833571}.
Similar concerns have also been documented in the grey literature, highlighting both the increased adoption as well as the generation of insecure code~\cite{snykAIsec}.
Therefore, the manual validation will allow us to evaluate the reliability of using LLMs for detecting and refactoring misconfigurations.

To the best of our knowledge, prior research did not propose or investigate the use of APR or LLMs to detect and repair Helm charts.

\section{Research Questions}
\label{sec:research_questions}

The goal of this research is threefold: investigate what misconfigurations are reported by existing tools, evaluate LLMs' refactoring capabilities in removing misconfigurations, and finally, manually measure the robustness of both tools and LLMs in terms of misconfigurations true positives and false negatives.

\finding{
\textbf{\textit{Overarching Hypothesis.}} 
Because most tools are rule-based, we hypothesize that they will always raise an alert for the same misconfiguration regardless of the Helm chart.
}

For example, a privileged container will always be detected, despite whether it is defined within a StatefulSet, a Deployment, or a Pod.
We plan to confirm this hypothesis by analyzing all (thousands of) charts available on Artifact Hub.

Specifically, we first aim to systematically evaluate a set of open-source analyzer tools on all Helm charts to measure how many misconfigurations are reported, of what type, whether there are certain misconfigurations reported by only a subset of tools (i.e., unique policies), and which is the tool with the best coverage (i.e., that finds most misconfigurations).

\begin{enumerate}
    \item[]\hspace*{-4ex}\textbf{RQ1.} \textit{What are the misconfigurations reported in Helm charts on Artifact Hub?}
\end{enumerate}

Second, automatically suggesting how to remove misconfigurations is still an open question, as it may depend on the domain of the application, and a fixed set of mitigations may not be an appropriate solution.
Therefore, it is interesting to investigate whether LLMs can be used to refactor charts by removing such misconfigurations. 

\begin{enumerate}
    \item[]\hspace*{-4ex}\textbf{RQ2.} \textit{Can LLMs refactor Helm charts and remove misconfigurations?}
\end{enumerate}

To this aim, we will query LLMs to refactor the chart for each misconfiguration detected in the previous step.
However, LLMs can hallucinate, i.e., generate wrong and inconsistent results, because, for example, of flawed training data sources~\cite{huang2023survey}.
Therefore, because of such hallucination problems, LLMs can refactor a chart with both secure and insecure configurations.
For example, it might be that mitigations are correct only for very simple modifications (e.g. replacing an empty memory requirement with memory requirements with a constant value), but more complex suggestions might escape detection by the toolchain. 
Therefore, we want to further evaluate the mitigations implemented by LLMs.

\begin{enumerate}
    \item[]\hspace*{-4ex}\textbf{RQ3.} \textit{Are there false positives in the analyzer tool results and hallucinations in LLM refactoring?}
\end{enumerate}

We will answer this question in two steps.
First, we will run the tools again on the chart refactored by an LLM and compare the output with the results of RQ1.
In other words, we will measure how many misconfigurations implemented by the LLM satisfy each tool policy.
Furthermore, because LLM can hallucinate, or analyzer tools can produce false positive results, we will perform a manual validation on a subset of charts to check whether misconfigurations are actually present or not and whether the implemented mitigations are logically correct.

For each RQ, we will collect several metrics and evaluate each tool individually using statistical tests. 
For RQ3, we will also perform a manual validation on a sub-sample of the results to provide the Agresti-Coull-Wilson interval for the reliability of our estimates. 
This interval will allow us to extrapolate the result of our manual validation on the dataset subsample and extend the conclusion to the all dataset.
The next section discusses what metrics and tests will be used.

\section{Execution Plan}
\label{sec:execution_plan}

\begin{figure*}[tb]
    \centering
    \includegraphics[width=0.9\textwidth]{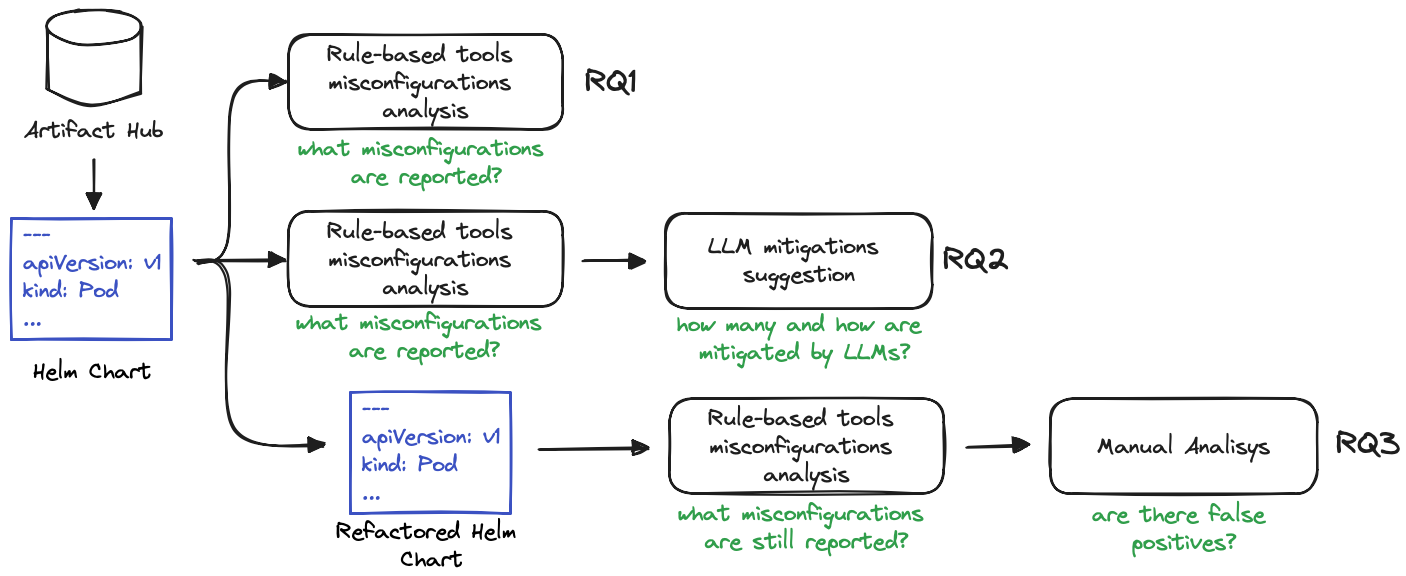}

    
    \begin{minipage}{0.9\linewidth}{\itshape \small
    To answer \textbf{RQ1}, we will analyze each chart with rule-based tools (such as Checkov and Datree), and measure what type and how many misconfigurations are reported.
    For \textbf{RQ2}, we will do the same but then, we will query an LLM (e.g., ChatGPT or Gemini) to refactor the chart to remove each misconfiguration reported at the previous step; for each misconfiguration, we will evaluate whether the LLM provided a correct or wrong fix, or refused to answer.
    Finally, for \textbf{RQ3}, we will perform a results validation by running the rule-based tools on the refactored charts and by manually analyzing (e.g., for false positives) a subset of charts.
    }\end{minipage}
    
    \caption{The execution plan of our experiment.}
    \label{fig:protocol}
    
\end{figure*}

In this section, we outline our approach to answering our research questions.
Figure~\ref{fig:protocol} summarizes the execution plan, where each step is further explained in the following subsections.

\subsection{Helm Charts Dataset}

Using the Artifact Hub APIs, we plan to mine all available charts on the repository, and locally save them; at the time of writing, there are $13.612$ charts available.
As an example, Table~\ref{tab:helm_chart} provides the list of the ten most popular Helm Charts available on Artifact Hub. For each chart, it provides the name, repository (to uniquely identify a chart), number of stars (used to rate the chart's popularity), the number of YAML chart template file lines, and the number of containers.

{\renewcommand{\arraystretch}{1.2} 
\begin{table}[t]
    \caption{Top ten most popular Helm charts in Artifact Hub.} 
    \label{tab:helm_chart}
    \centering
    \begin{minipage}{\linewidth}{\itshape \small These are the ten most popular Helm Charts on Artifact Hub, ordered by the number of stars.
    For each chart, it shows the corresponding name, repository (used to distinguish charts with the same name), number of stars (used to rate popularity), and number of lines in the chart's YAML template generated with the \texttt{helm template} command, and the containers deployed by the chart. 
    }\end{minipage}
    
\begin{tabular}{p{2cm} p{1.5cm} c c c}
    \toprule
    \centering \thead{Name} & 
    \centering \thead{Repository} & 
    \centering \thead{Stars} & 
    \centering \thead{Lines} & 
    \centering \thead{\#containers} \tabularnewline

    \midrule
    kube-prometheus-stack & prometheus-community & $572$ & $44,331$ & $11$ \\

    \midrule
    ingress-nginx & ingress-nginx & $465$ & $725$ & $3$ \\

    \midrule
    cert-manager & cert-manager & $446$ & $1,313$ & $4$ \\
    
    \midrule
    argo-cd & argo & $325$ & $18,539$ & $9$ \\
    
    \midrule
    prometheus & prometheus-community & $297$ & $1,278$ & $6$ \\
    
    \midrule
    redis & bitnami & $269$ & $645$ & $2$ \\
    
    \midrule
    grafana & grafana & $258$ & $332$ & $2$ \\
    
    \midrule
    kubernetes-dashboard & k8s-dashboard & $220$ & $532$ & $1$ \\
    
    \midrule
    postgresql & bitnami & $218$ & $225$ & $1$ \\
    
    \midrule
    traefik & traefik & $175$ & $265$ & $1$ \\

    \bottomrule
    
    \end{tabular}
\end{table}
}

From all Helm charts retrieved from Artifact Hub, we will remove duplicates, charts with YAML syntax errors, and charts that all tools will fail to analyze (e.g., because the file size is too large or because of missing information).
As long as the tools can analyze the chart YAML template, we will not discard a chart based on the size.

\subsection{Analyzer Tools}

We retrieved open-source Helm Charts and YAML file analyzers either from the CNCF projects list or from GitHub, listed in Table~\ref{tab:static_tools}.
We will use these tools to analyze each chart and answer our research questions.
{\renewcommand{\arraystretch}{1}
\begin{table}[t]
    \caption{Open-source Helm Charts and YAML files analyzers.}
    \label{tab:static_tools}
    \centering
    \small
    \begin{minipage}{\linewidth}{\itshape 
        All tools are open-source static-time analyzers that can check Helm Charts and K8s manifests.
        All tools are part of the CNCF as graduated or incubating projects, except Kube-linter and Kubeaudit, which were retrieved from GitHub.
        We did not consider PodSecurityPolicy, because it was officially deprecated in K8s v1.21, and infrastructure policies (e.g., "\textit{Ensure that the --kubelet-https argument is set to true}") because it is not possible to check them by analyzing Helm Charts.
    }\end{minipage}

    \begin{tabular}{llcrcl}
    \toprule
    \centering \thead{Tool}
    & \thead{Company}
    & \thead{CNCF}
    & \thead{Policies}
    & \thead{Fix}
    & \thead{Link} \tabularnewline

    \midrule
    Checkov & 
    BridgeCrew & 
    \ding{51} &
    \centering $38$ & 
    & 
    \cite{checkov} \\

    \midrule
    Datree &
    Datree & 
    \ding{51} &
    \centering $38$ &
    &
    \cite{datree} \\

    \midrule
    KICS &
    Checkmarx & 
    \ding{51} &
    \centering $31$ &
    &
    \cite{kics} \\

    \midrule
    Kube-linter &
    StackRox & 
    &
    \centering $51$ &
    &
    \cite{kubelinter} \\

    \midrule
    Kubeaudit &
    Shopify & 
    &
    \centering $14$ &
    \ding{51} & 
    \cite{kubeaudit} \\

    \midrule
    Kubescape &
    ARMO & 
    \ding{51} &
    \centering $50$ &
    & 
    \cite{kubescape} \\
    
    \midrule
    Terrascan &
    Tenable & 
    \ding{51} &
    \centering $33$ &
    &
    \cite{terrascan} \\
    
    \bottomrule
\end{tabular} 
\end{table}
}

We selected the tools that are part of the CNCF as ``graduated'' or ``incubating'' projects.
In addition, we included Kube-linter and Kubeaudit to further increase the sample of existing tools (see Table~\ref{tab:static_tools}).
Each tool supports different policies to check for misconfigurations in K8s manifests.
Example policies include ensuring that each container has a configured memory limit, that container CPU requests are not equal to its limits, minimizing the admission of containers with the \texttt{SYS\_ADMIN} capability, and ensuring that each container image has a pinned (tag) version.
Also, all tools allow defining custom policies to ignore specific resources, specify whether to scan a Helm Chart folder, a set, or a single YAML file, and provide different output formats from the console to the JSON format.


\subsection{LLM Query Engineering}

To implement mitigations, we will use two LLM chatbots, specifically, chatGPT~\cite{chatgpt} and Google Gemini~\cite{gemini}.
Because of the limited input size that LLM chatbots can parse, and because Helm chart templates can contain up to thousands of lines, we can not provide for input the whole chart.
However, misconfigurations affect either one line or a single YAML object, which is reported by existing tools along with the violated policy.

Therefore, to answer both RQ2 and RQ3, we will query LLM chatbots to refactor portions of Helm charts to remove the misconfigurations that were detected in the previous step.

Each query will include i) a K8s resource, ii) a misconfiguration, and iii) the resource YAML code; the second part of the query will specify the format of the chatbot output for further processing.

For example, for a Deployment object and the ``\textit{Prevent containers from escalating privileges}'' policy~\footnote{\url{https://hub.datree.io/built-in-rules/prevent-escalating-privileges}}, the query will be the following:

\smallbreak
\noindent
\textit{Refactor the following Deployment K8s resource to prevent containers from escalating privileges. Output only the refactored YAML file. [YAML code ...]}

\textbf{Example on Listing~\ref{listing:misconfig} and Datree tool.} Listing~\ref{listing:misconfig} shows an example of a busybox Pod without resources memory requests defined. 
Datree will raise an alert when analyzing this resource, highlighting that memory requests should be defined.

\begin{lstlisting}[language=xml,caption={An example of a Helm chart memory requests misconfigurations.\label{listing:misconfig}}, basicstyle=\footnotesize,breaklines=true,numbers=left,linewidth=\columnwidth] 
...
---
apiVersion: v1
kind: Pod
metadata:
  name: busybox-pod
  namespace: busybox-namespace
spec:
  containers:
  - name: busybox-container
    image: busybox:1.36
    imagePullPolicy: Always
    resources:    
      requests:
        cpu: 250m
    ports: ...
---
...
\end{lstlisting}

To refactor this resource, we can give as input the K8s resource, i.e., Busybox pod, and the violated policy, i.e., ``\textit{Ensure each container has a configured memory request}''~\footnote{\url{https://hub.datree.io/built-in-rules/ensure-memory-request}} to LLM chatbots, along with the Pod YAML code shown in Listing~\ref{listing:misconfig}, as follows:
%

\smallbreak
\noindent
\textit{Refactor the following Pod K8s resource to ensure each container has a configured memory request. Output only the refactored YAML file.} 

Listing~\ref{listing:fix} shows the refactored Pod YAML by Google Gemini, using the previous query. When this refactored chart is checked by Datree again, no memory limit alert is raised, confirming that the generated suggestion by Google Bart satisfies the memory requests policy.

\begin{lstlisting}[language=xml,caption={An example of a Helm chart refactoring by Google Gemini.\label{listing:fix}}, basicstyle=\footnotesize,breaklines=true,numbers=left,linewidth=\columnwidth]
...
    resources:    
      requests:
        memory: 250Mi
        cpu: 250m
...
\end{lstlisting}

Because every tool, for each failed policy, will output a text description of the corresponding policy (e.g., ''\textit{Ensure each container has a configured memory request}``), we will use this description to build the query for the LLM.
This will allow us to automatically craft all the queries from each tool output; during our preliminary study, using such descriptions to query the LLMs proved to be successful in implementing mitigations.
Also, because for equivalent policies of different tools, the description can be slightly different, this will allow us to evaluate whether a different description yields better mitigation or not, for the same misconfiguration.


\subsection{LLMs Interaction Process}

To implement the proposed pipeline to automatically find misconfigurations and use LLMs for remediation, we will use a Python program and LLM APIs.
Specifically, at the beginning, one of the tools will analyze a chart and output the results in YAML format; for each tool and chart, the result YAML file will include a unique identifier to the failed policy (e.g., container without memory limits) and the resource to which the misconfigurations refers to (i.e., resource type, name, and namespace).
Therefore, the Python program will first parse the tool analyzer output and for each reported misconfiguration, it will locate and extract the YAML snippet of the affected resource from the chart.
Using the policy text description and the resource YAML snippet, the program will then query the LLM using the available APIs (e.g., Python SDK for the Gemini API~\footnote{\url{https://ai.google.dev/tutorials/python_quickstart}}), asking to refactor the resource to remove the misconfigurations.
While querying the LLM, the program will also ask to only output the refactored YAML file, such that it can be locally saved and compared to the original YAML snippet.
In the end, after saving the LLM output, the program will compare with a diff the original YAML snippet with the LLM output YAML snippet to check whether mitigation was implemented or not.
For the charts and misconfigurations refactored by the LLM, we will run the analyzer tools again and perform manual analysis on a subset of chart to check whether the mitigations satisfy the tool policies and are logically correct (e.g., memory is not equal to $0$).


\subsection{Analysis}

For each RQ, we explain what are the collected metrics and how we plan to analyze the collected data to answer each question.

For \textbf{RQ1}, we will collect the number of detected misconfigurations, and the number of unique detected misconfigurations by each tool.
We will configure the tools to output a JSON, parse the results, and count each misconfiguration ID.
For every tool output, we report the number of resources for which a policy failure was reported, and the top violated policies.

For \textbf{RQ2}, in the first step, we will collect the same metrics as in RQ1.
Then, to measure the LLM refactoring, we will measure as a \textit{proportion} of the total failed policies per tool:

\begin{itemize}
    \item correct refactoring;
    \item wrong refactoring;
    \item refused refactoring (e.g., the LLM did not answer).
\end{itemize}

To evaluate such three cases, we will run the tools again on the refactored templates and, for each misconfiguration, measure:

\begin{itemize}
    
    \item a correct refactoring, if there is a \texttt{diff} between the original and refactored templates, and the tool does not output an alert anymore;

    \item a wrong refactoring if there is a \texttt{diff} between the original and refactored templates, and the tool still outputs an alert (i.e., LLM changed the configuration in a way that is not accepted by the tool, thus we consider it wrong).

    \item a refused refactoring if there is no \texttt{diff} between the original and refactored templates (i.e., LLM did not change the chart); obviously, the tool should also still raise an alert.
    
\end{itemize}

At the end of this process, we will have the numbers of mitigations that were correctly, wrongly, and not mitigated by an LLM.
To automatize this process, we will output each tool results in a JOSN format, that can be easily parsed, and also query the LLM such that the output is only a YAML file.

Finally, for \textbf{RQ3}, we will perform both automatic and manual validation.
In particular, the automatic validation will consist of running the analyzer tool on the charts refactored by the LLM to evaluate whether the mitigations, if implemented, satisfy the tool's policies or not.
At the same time, on a subset of charts, we will perform a manual validation consisting of verifying both the tool output, specifically, whether there are false positives, and the mitigations, whether they are logically correct. 
Because the manual validation will be performed on a subset of charts, we will compute the Agresti-Coull-Wilson interval, which is a method to construct an accurate and reliable confidence interval, also for small sample sizes. 
This interval will allow us to extrapolate the results, in terms of false positives, from a chart subset, instead of analyzing all thousands of charts.

Table~\ref{tab:metrics} shows, for each RQ, the collected metrics during the execution of the experiment.

{\renewcommand{\arraystretch}{1} 
\begin{table}[t]
    \caption{Recorded metrics per RQ.}
    \label{tab:metrics}
    \centering
    \small

    \begin{tabular}{llp{5cm}}
    \toprule
    \centering \thead{RQ} 
    & \thead{Metric} 
    & \thead{Description} \tabularnewline
    
    \midrule
    \multirow{2}{*}{RQ1} & N\_MISC & Number of misconfigurations reported per tool. \\
    & U\_POL & Type and number of uniquely reported misconfigurations per tool. \\

    \midrule
    \multirow{5}{*}{RQ2} & N\_MISC & Number of misconfigurations reported per tool. \\
    & LLM\_C & Type and number of correct refactorings.\\
    & LLM\_W &  Type and number of wrong refactorings.\\
    & LLM\_R &  Type and number of refused refactorings.\\

    \midrule
    \multirow{3}{*}{RQ3} & T\_TP &  Type and number of tool true positives. \\
    & T\_FP & Type and number of tool false positives. \\
    & LLM\_I\_C & Type and number of correct refactorings.\\
    & LLM\_I\_W & Type and number of wrong refactorings.\\
    & LLM\_I\_R & Type and number of refused refactorings.\\

    \bottomrule
\end{tabular}
\end{table}

\section{Preliminary Study}
\label{sec:preliminary_study}

As a preliminary study, we retrieved $60$ publicly available Helm Charts from Artifact Hub, ordered by popularity, and analyzed them with the seven open source tools shown in Table~\ref{tab:static_tools}.
Besides computing descriptive statistics, such as the most common misconfigurations found so far, which are too permissive ClusterRoles, missing memory limits for resources, and using the default namespace, we also found interesting preliminary findings.

An interesting issue common to all tools is the definition of what a good fix is. For example, a policy common to every tool is to check whether CPU and memory limits are defined for every resource. However, assigning incorrect values as limits (i.e., $0$, very large values, or the string ``john'') is sufficient to satisfy the policy.
Similarly, we found interesting false negative results, i.e., configurations that satisfy tool policies but have no effect in practice.
For example, Checkov and KICS raise an issue if there is a deployment-like resource (e.g., Pod, Deployment, or StatefulSet) in the chart template that is not bound to any network policy.
However, Checkov only checks whether a network policy object is defined, but not if the policy is within the same namespace as the resource, or if there are any network ingress and egress rules; therefore, defining an empty network policy is sufficient to pass the check, even though will have no effect in practice.
The previous examples highlight the biggest problem of using several tools, that is, a lack of policy standardization.
Indeed, even if a policy is equivalent between different tools (e.g., defining a network policy), tools may not check the same set of resources for the same policy (e.g., only Pod and Deployment, but not Service), or can accept different mitigation compared to another tool (e.g., only ingress rules instead of both ingress and egress rules).

We also found the presence of false positives in most tool outputs, i.e., alerts were raised even if a misconfiguration was not present.
For example, to prevent containers from accessing the underlying host, Datree checks whether deployment-like resources mount volumes with the \texttt{hostPath} key (e.g., \texttt{hostPath: path: /proc}), eventually raising the corresponding issue.
However, if the \texttt{volumes} list is empty or null, Datree still raises an issue, even if there is no misconfiguration.

Based on this preliminary study findings, further research is needed to shed light on these problems.

\section{Threats to Validity}
\label{sec:threats_to_validity}

\textit{Internal Validity}. We will mine all publicly available charts on Artifact Hub, however, private charts or charts deployed by companies in private clusters might yield different results.

\textit{External Validity}. To evaluate each chart, we only consider a set of seven open-source chart analyzer tools, however, other tools (such as custom or closed-source tools) might also yield different results.
Also, refactored configurations by LLMs may depend on the provided queries; therefore, different queries can generate different results.

\textit{Construct Validity}. We assume the chart analyzer tools have not been compromised and that the results are correct; however, eventual errors or bugs might also yield different results.


\begin{acks}
This work was partly supported by the European Commission under grant n.101120393 (Sec4AI4Sec), and the Dutch Research Council (NWO) under grant n.NWA-1215.18.006 (Theseus) and n.KICH1.VE01.20.004 (HEWSTI).
\end{acks}

\subsection*{CRediT statements}
	\emph{Conceptualization:} Francesco Minna, Fabio Massacci, Katja Tuma; 
	\emph{Methodology:} Francesco Minna, Fabio Massacci, Katja Tuma; 	
	\emph{Software:} Francesco Minna (initial); 	
	\emph{Validation:} not yet;	
    \emph{Formal analysis:} not yet;	
    \emph{Investigation:} Francesco Minna (preliminary);	
    \emph{Resources:} Katja Tuma;	
    \emph{Data Curation:} not yet; 	
    \emph{Writing - Original Draft:} Francesco Minna
    \emph{Writing - Review \& Editing:} Francesco Minna, Fabio Massacci, Katja Tuma 
    \emph{Visualization:} not yet 
    \emph{Supervision:} Fabio Massacci, Katja Tuma 
    \emph{Project administration:} Fabio Massacci, Katja Tuma	
    \emph{Funding acquisition:} Fabio Massacci, Katja Tuma 

\clearpage

\bibliographystyle{ACM-Reference-Format}

\end{document}